\documentclass[twocolumn]{aastex63}

\newcommand{\sgra}{$\rm Sgr~A^{\star}$}

\received{ }
\revised{ }
\accepted{\today}
\submitjournal{ApJ}

\shorttitle{star cluster in a tidal field}
\shortauthors{Ivanov and Lin}

\renewcommand{\textcolor}[2]{#2}

\begin{document}

\title{The secular evolution of a uniform density star cluster immersed in a compressible galactic tidal field}

\correspondingauthor{}
\email{pbi20@cam.ac.uk}

\author{P. B. Ivanov}
\affil{ Astro Space Centre of P. N. Lebedev Physical Institute,  84/32 Profsoyuznaya st., Moscow, GSP-7, 117997, Russia}

\author{D. N. C. Lin}
\affiliation{Department of Astronomy and Astrophysics, University of California, Santa Cruz, USA}
\affiliation{Institute for Advanced Studies, Tsinghua University, Beijing, 100086, China}

\begin{abstract}
Nuclear stellar clusters are common in the center of galaxies.  We consider the possibility
that their progenitors \textcolor{assumed to be globular clusters}{}may have formed elsewhere, migrated to and assembled near their present 
location.  The main challenge for this scenario is whether globular clusters can withstand the tidal
field of their host galaxies. Our analysis suggests that provided the mass-density distribution 
of background potential is relatively shallow,  as in some galaxies with relatively flat surface
brightness profiles, the tidal field near the center of galaxies may be shown to be able to compress 
rather than disrupt a globular cluster at a distance from the center much smaller than the conventionally 
defined `tidal disruption radius', $r_t$. To do so, we adopt a previously constructed formalism and consider the secular 
evolution of star clusters with a homogeneous mass density  distribution.  We analytically solve the secular 
equations in the limit that the mass density of stars in the galactic center approaches a uniform distribution.  
Our model indicates that a star cluster could travel to distances much smaller than $r_t$ without disruption, 
thus potentially contributing to the formation of the nuclear cluster. However, appropriate numerical N-body 
simulations are needed to confirm our analytic findings.

\end{abstract}


\section{Introduction} \label{sec:intro}

GAIA data reveal the prevalence of stellar streams in the Galaxy
\citep{Myeong2018, Myeong2019, Helmi2018, Koppelman2018, Deason2018, Necib2019}.  
Similar structures are also found in M31 \citep{Guhathakurta2006, Gilbert2009}.  
They are thought to be the debris of tidally disrupted stellar clusters
or dwarf galaxies\citep{Johnston1995}. In the context of the $\Lambda$CDM 
scenario of galaxy formation, their progenitors are building blocks
which converge to form larger galaxies surrounded by dark matter potential
\citep{WhiteRees1978, Blumenthal1984, Davis1985, Navarro1995, Navarro1996, 
Navarro1997}. 
Along the course of their coalescence, loosely bound sub-structures are 
subjected to tidal disruption \citep{Ibata1994, Oh1995, Newberg2016} and 
their debris streams contribute to the 
dynamical structure of the merger byproducts\citep{LyndenBell1995}.  Some 
compact systems may withstand the tidal perturbation due to the galactic 
potential and be retained as globular clusters\citep{Fall1977}.   
The conventional stability boundary is the ``tidal disruption radius'' $r_t$ 
where the intruding or satellite systems' average mass density is comparable 
to that contribute to the galactic potential. If these systems can preserve 
their integrity on their way to the central regions 
of galactic conglomerates, they could also lead to the development of cusps 
versus cores\citep{Tremaine1975, Tremaine1976a, DekelDevor2003, DekelArad2003}.  

Today, there are several stellar clusters, including the Archies and Quintuplet
clusters, in the vicinity of the Galactic Center\citep{Nagata1995, Cotera1996, 
Kobayashi1983}.  These clusters have much higher 
internal density and contain many more massive stars than all the known globular 
clusters in the Galaxy\citep{Figer1999, Espinoza2009}. Ideally, these clusters 
could have undergone orbital 
decay from a few kpc away to their present location within the Hubble time\citep{Gerhard2001}.  
But, the brief (a few Myr) lifespan of massive main sequence stars contained in them 
casts a strong limit on the distance over which they may have migrated.  
Moreover, the intense tidal perturbation by the Galactic potential poses a challenge 
to their protracted sustainability \citep{PortegiesZwart2002, Gurkan2005}. Based on these considerations, it has
been suggested that these clusters were formed close to their present-day location
\citep{Figer2002}.

Within 1 pc from the very center of the Galaxy, a nuclear cluster with $\gtrsim 10^7$ 
stars surrounds a $M_{SMBH} \simeq 4.2 \times 10^6 M_\odot$ black hole commonly dubbed
as \sgra \citep{Ghez1998, Genzel1997, Genzel2010}. 
Although stars in the nuclear cluster are predominantly low-mass and 
mature\citep{Do2009}, there is a population of young OB and Wolf Rayet 
stars \citep{Ghez2003}. While the young stars may be formed \citep{Goodman2003, Levin2003, 
Nayakshin2007} or rejuvenated 
\citep{Artymowicz1993} {\it in situ} within the past few Myr, the old star could have migrated to this
confined central region if they were once members of some progenitor clusters which 
preserved their dynamical integrity during the course of their orbital evolution
\citep{Gerhard2001, Madigan2014}.

Many nucleated dwarf galaxies are found in the central regions of galaxy clusters
\citep{Sandage1984}.  Nuclear clusters are also commonly found in other Milky Way-type
disk galaxies\citep{KormendyHo2013}.  Their contribution to the surface brightness 
distribution is conspicuous in their S{\'e}rsic profiles \citep{Boeker2002, misgeld2011}.  At the center 
of the more massive early-type elliptical galaxies, unresolved point-sources 
of light are ubiquitous \citep{Ferrarese2005} without significant, if any, 
contribution from nuclear clusters.  Highly variable sources which outshine 
their host galaxies over multi-wavelength are dubbed as active galactic nuclei (AGNs).  
They are thought to be powered by disk accretion onto massive black holes 
\citep{LyndenBell1969}. 

The luminosity of the nuclear clusters and massive black holes can be distinguished 
from that of the host-galaxy background through the decomposition of the S{\'e}rsic 
and cusp photometric surface brightness distribution. The velocity dispersion of 
the host galaxies' bulge can be obtained independently with spectroscopic 
measurements.  Despite the dichotomy between the mass and morphological classification 
of their host galaxies, surveys indicate that the mass $M_c$ of both nuclear clusters 
and massive black holes are correlated with the velocity dispersion $\sigma$ in the 
bulge of their host galaxies\citep{Gebhardt2000, Ferrarese2000}.  They 
have similar power-law $M_c-\sigma$ relationships \citep{Ferrarese2006}  albeit 
for intermediate-mass galaxies (such as the Milky Way) which contain both population, 
the nuclear clusters are on average a few times more massive than the massive 
black holes\citep{KormendyHo2013}.   

These tantalizing general scaling laws signal the possibility of some links between the 
dynamics of nuclear clusters during the evolution from relatively low-mass to massive
galaxies.  If the merger tree is the pathway for galactic assembly, nuclear clusters 
and central massive black holes would coagulate together with the host building block
galaxies\citep{Pfeffer2014}.  After their orbits are virialized, relatively massive entities may undergo 
further orbital decay due to the effect of dynamical friction\citep{Tremaine1976b, Just2011, neumayer2020}. One 
important issue is under what condition can dense stellar clusters survive tidal 
disruption on their way to the center of galactic bulge\citep{vanderMarel2007, fellhauer2007}.

The smallest and most common dwarf galaxies represent microcosm of such evolutionary
pathway\citep{Ferguson1991}.  Some dwarf galaxies contain multiple globular clusters.  
For example, the nearby Fornax dwarf spheroidal galaxy (dSph) hosts six globular clusters 
\citep{wangfornax2019} and their orbital
decay time scale, due to dynamical friction, has been estimated to be less than 1 Gyr
\citep{Hernandez1998}.  These clusters remain in the field of Fornax due to the tidal 
stirring by the Galactic halo potential\citep{Oh2000a}.  
In contrast, many nucleated dwarf galaxies are found inside the much larger core radius 
(on Mpc scales) of some galaxy clusters\citep{Binggeli1991}. These nucleated dwarfs 
are characterized by central cusps in their surface brightness distribution.  
Moreover, some of these nucleated dwarfs also nest globular clusters \citep{Miller2007}.  Their
nucleated structure, including that of ultra-compact dwarf galaxies\citep{Drinkwater2003} 
may be byproducts of merging globular clusters\citep{Goerdt2008}.  In order to account for the
dichotomy between multiple floating globular clusters in the Fornax dSph and the omnipresence
of nucleated dwarf galaxies in the central cores of galaxy clusters, \citet{Oh2000b} suggest 
that the tidal perturbation from the cluster of galaxies is compressive due to its shallow 
density slopes \citep{Navarro1996}. Similar process may also play a role in the formation
of the S{\'e}rsic surface brightness profile found in most elliptical galaxies\citep{Emsellem2008},
heating of disk galaxies in the center of galaxy clusters\citep{Valluri1993}, and globular
clusters during their crossing of Galactic disk\citep{Ostriker1972}.

As dwarf galaxies coalesce into larger entities, nuclear clusters on 
different branches of the merger tree also converge.  After the post-merger 
virialization, the nuclear clusters' ability to undergo orbital decay 
and to survive against tidal disruption determine the $M_c$ and $\sigma$ 
values for their amalgamated byproducts.  The accumulation of multiple nuclear
clusters in confined regions may also promote the emergence of massive 
black holes\citep{CapuzzoDolcetta1993}.  Finally, pre-existing massive black holes 
in the center of elliptical galaxies may maintain their local dominance by
tidally disrupting incoming nuclear clusters\citep{Gerhard2001} at $r_t$ 
comparable to or larger than the massive black holes' radius of dynamical 
influence (i.e. $\sim G M_{SMBH}/ \sigma^2$). Similarly
newly arriving massive black holes may also disrupt pre-existing nuclear clusters.  
This effect may account for the exclusion of nuclear clusters around massive black holes
in the center of elliptical galaxies.
 
In galaxies with highly peaked central mass concentration and 
steep declining surface brightness gradient, the critical condition for tidal 
disruption of a globular cluster is similar to that of stars around SMBH or
planets around stars.  An entity with a mass $M_0$, radius $R_0$, an average density 
$\rho= 3 M_0/4 \pi R_0$, and a parabolic orbit undergo tidal disruption around 
a point mass $M_G$ when their periastron distance between them is smaller than 
a few times the tidal disruption radius $r_t = (M_G/\rho)^{1/3}$or equivalently when
the ``average density" associated with the point-mass potential $\rho_G = 3 M_G/4 
\pi r_t^3$ is $\gtrsim \rho$\citep{FrankRees1976}.  A similar tidal limiting 
radius also applies to self gravitating entities on a circular orbit
\citep{Chandrasekhar1969}.  But around the central regions of some galaxies 
where the density is a weakly declining function of 
distance from them, this condition is modified by the additional gravity 
from the background stars in the concentric shells which sandwich the satellite system.
Qualitatively, around a homogeneous background, stars further away from 
the center of the bulge accelerate more rapidly than those closer to the
center.  This effect leads to a tidal compression\citep{Oh2000b, Masi2007}.  

In this paper, we provide a quantitative analysis to verify the possibility that a compressive 
than disruptive tidal field could preserve integrity of globular clusters orbiting around a spherically 
symmetric distribution of mass at distances much smaller than $r_t$.  
In \S2, we consider an idealized analytic  model to examine the condition for tidal stability of a stellar cluster 
following the work of \cite{MitchellHeggie2007}, which is based itself on the model of so-called 
\citet{Freeman1966a, Freeman1966b, Freeman1966c} bar. This model has the advantage that 
the cluster immersed in a stationary tidal field maintains uniform distribution of its mass
density, $\rho$, and has the shape of a general ellipsoid with unequal semi-major axes. 
This approach greatly simplifies analytic analysis of the model. Then, we formulate
equations describing secular evolution of the model
proceeding when its orbit assumed to be circular shrinks as a result of dynamical friction. In \S3, 
we discuss solutions to the secular equations. These solutions describe the adiabatic
adjustment in the phase space distribution subjected to changes in the external tidal fields.
At first, we consider the strong tidal limit and determine the critical tidal 
disruption condition for power-law density distribution for the background galaxy, $\rho_G \propto R^{-k}$ 
under the assumption  that the density in the galactic background decreases with radius gradually, and, accordingly, $k$ is small.
We show that in this case, as expected in the bulge of some galaxies,  these clusters are practically indestructible by the
tidal perturbation of the background galaxy.   Later in this section we show that 
the cluster in our model remain spherically symmetric in the formal limit $k=0$ corresponding to the homogeneous density distribution for any strength of
tidal field and that its radius, $a$, is described by a solution to a quartic equation. We take into account a non-zero, but formally small value of $k$ in the framework of a perturbation theory and show how the critical semi-major axes of the cluster as well as its density depends on the strength of the tidal field. We note that clusters with centrally concentrated 
density profile are more likely to survive tidal disruption than the homogeneous model we have adopted. Therefore our criteria for
clusters' preservation in a relatively shallow background potential is robust.
We explore some astrophysical application based on the King model, several commonly used 
parameterized models,  the empirical S{\'e}rsic model for 
galactic bulges and elliptical galaxies, and composite model for the Milky Way galaxy in \S4.  We summarize our results and discuss
their limitations and implication in \S5. Additionally, in \S5 
we provide a qualitative argument, which allows us to suggest
that at least some more realistic models of star cluster 
evolution in a tidal field of a galaxy corresponding to nearly
homogeneous mass density of galactic matter could behave 
similarly to our idealized toy model.
  
\section{An analytic model of a star cluster in compressive tidal field}

For mathematical convenience, we adopt the boundary conditions that: 1) the density
of the background galaxy is spherically symmetric, 2) the cluster is on a circular 
orbit around the center of the galaxy, and 3) the mass density inside the cluster 
is homogeneous.  Under these conditions, we 
consider 1) the gravitation potential 
in terms of a triaxial ellipsoid (\S2.1), 2) the solutions of the equation of motion 
for stars in the frame which is comoving with the cluster and corotate with its orbital frequency (\S2.2), and 3)  
normal modes, frequencies of stellar motion, and adiabatic 
invariants in terms of action variables associated with the normal modes.  These quantities enable 
us to extrapolate the density and shape adjustments to gradual increase in the
tidal potential (\S2.3).  Physically, this approximation represents the slow decay of 
the cluster's nearly circular orbit to the proximity of the galactic center, 
starting from very large galactic distances where the external field is negligible 
and the stellar cluster is spherical symmetric.  

\subsection{Basic definitions  and relations} \label{sec:style}
\label{basic}
We adopt the same non-inertial right handed Cartesian 
coordinate system as in \cite{BertinVarri2008} with $x$, $y$ and $z$ axes
directed outward galactic centre, in the orbital plane and perpendicular to it, respectively. In this system
equations of motion of stars take the form
\begin{equation}
\ddot x-2\Omega \dot y +{\partial \Phi \over \partial x}-\gamma^2 x=0, \quad \ddot y+2\Omega \dot x 
+{\partial \Phi\over \partial y}=0, \quad \ddot z+{\partial \Phi\over \partial z}+\Omega^2 z=0,
\nonumber
\label{e1}
\end{equation} 
where dot stand for time derivative, $\Omega$ is angular frequency of orbital motion assumed to be circular:
\begin{equation}
\Omega^2={1\over R}{\partial \over \partial R}\Phi_{G}, 
\label{e2}
\end{equation} 
$R$ is the distance from galactic centre and $\Phi_{G}$ is gravitational potential of a galaxy. The quantity
$\gamma$ can be expressed in term of $\Omega$ and epicyclic frequency, $\kappa$, as $\gamma^2=4\Omega^2-\kappa^2=
{1\over R}{d\over dR}\Phi_G-{d^2\over dR^2}\Phi_G$. For a spherically symmetric distribution of galactic mass density, $\rho_{G}$, assumed from now on we can
express $\Omega$ and $\gamma$ in terms of $\rho_{G}$ as
\begin{equation}
\Omega^{2}={4\pi G\over R^{3}}\int R^2dR \rho_G, \quad  \gamma^2=-{4\pi G\over R^3}\int R^{3}dR {d\over dR}\rho_{G},
\label{e3}
\end{equation} 
where $G$ is gravitational constant. Note that the latter equation yields $\gamma^2 > 0$. 

Physically, the sign of $\gamma^2$ is
determined by interplay between tidal and centrifugal
forces acting in $x$ direction, relative to the cluster 
center. Although it is easy to show that the tidal 
force is attractive, when $\gamma^2 < \Omega^2$ the 
centrifugal is always repulsive with its absolute 
value always larger than that of tidal force. Thus, 
the combination of two forces is always repulsive 
when $\gamma^2 > 0$, and neutral when $\gamma^2=0$,
which corresponds to a homogeneous density of galactic
stars.

The gravitational potential of stars in the cluster $\Phi$ obeys Poisson equation
\begin{equation}
\Delta \Phi =4\pi G \rho,
\label{e4}
\end{equation}
where $\Delta $ is Laplace operator and $\rho $ is mass density of the stars.

Equations (\ref{e1}) have the well known Jacobi integral
\begin{equation}
E={v^2\over 2}+\Phi_{ext} +\Phi,
\label{e5}
\end{equation}
where $v$ is the absolute value of velocity of a star,
\begin{equation}
\Phi_{ext}={\Omega^2 z^2\over 2}-{\gamma^2 x^2\over 2}
\label{enn5}
\end{equation} is the sum of potentials of tidal and centrifugal forces.

\subsection{Equations of motion in canonical form for a model with homogeneous density distribution}
\label{canonical}
In what follows we are going to use a model of a star cluster proposed in \cite{MitchellHeggie2007},
which  is  related  to  the \citet{Freeman1966a, Freeman1966b, Freeman1966c} models for uniform
density rotating bars.
Although the model is rather artificial it has the advantage that stellar density of 
the cluster is homogeneous and the cluster has the form of an ellipsoid. This allows for an analytic treatment of 
the problem on hand. 

We use below the fact that the gravitational potential of an ellipsoid having a uniform density $\rho$ has quadratic form
\begin{equation}
\Phi=\pi G \rho \sum_{i=1,3}A_ix_i^{2},
\label{e6} 
\end{equation}
where we set to zero unimportant constant part and the indices $1$, $2$ and $3$ stand for the $x$, $y$ and $z$,
respectively. The dimensionless quantities $A_i$ can be expressed in terms of two angles, $\theta $ and $\phi$ determined 
by ratios of $a_i$. Namely, let us arrange the axes $a_i$ in ascending order $a_{min}\le a_{int}\le a_{max}$ and introduce
$\theta $ and $\phi$ according to the relations  $\sin \theta =\sqrt {{a_{max}^2 -a_{int}^2\over a_{max}^2-a_{min}^2}}$ and 
$\cos \phi ={a_{min}\over a_{max}}$. It may be then shown that $A_i$ can be expressed in terms of incomplete elliptic 
integrals, see e.g. \cite{Chandrasekhar1969}. For our purposes it is, however, more convenient to use the equivalent explicit
expressions
\begin{equation}
A_{1}={2\cos \phi \Delta^{1/2}(\theta, \phi)\over \sin^{3}\phi}\int^{\phi}_{0}d\phi^{'}{\sin^{2} \phi^{'}\over 
 \Delta^{1/2}(\theta, \phi^{'})}, 
 \end{equation}
 \begin{equation}
 A_{2}={2\cos \phi \Delta^{1/2}(\theta, \phi)\over \sin^{3}\phi}\int^{\phi}_{0}d\phi^{'}{\sin^{2} \phi^{'}\over 
 \Delta^{3/2}(\theta, \phi^{'})},
\label{en6} 
\end{equation}  
where $\Delta(\theta,\phi)=1-\sin^2\theta \sin^{2}\phi$, and we take into account that $\sum_{i=1}^{i=3}A_i=2$ and, therefore
$A_3=2-(A_1+A_2)$. 

Using the expression (\ref{e6}) equations (\ref{e1}) can be brought in a standard form by introducing three
new frequencies 
\begin{equation}
\omega_{1}^2=2\pi G\rho A_1-\gamma^2, \quad \omega_{2}^2=2\pi G\rho A_2, \quad \omega^2_{3}=2\pi G\rho A_3+\Omega^2.
\label{e18}
\end{equation}
We have
\begin{equation}
\ddot x-2\Omega \dot y +\omega_1^2 x=0, \quad \ddot y+2\Omega \dot x +\omega_2^2y=0, \quad \ddot z+\omega_3^2 z=0,
\label{e19}
\end{equation} 
It is seen that motion in the vertical direction corresponds to a simple oscillator having the energy 
$E_3={1\over 2}({\dot z}^2+\omega_3^2z^2)$. It is well known that the so-called action variable 
\begin{equation}
I_3=E_3/\omega_3
\label{e20}
\end{equation}  
is an adiabatic invariant, which stays constant when parameters of the problem change slowly.

The 'horizontal' coordinates $x$ and $y$ are coupled by Coriolis force. Accordingly, motion in the horizontal 
direction corresponds to a two dimensional rotating oscillator. In order to introduce the action variables
$I_1$ and $I_2$ for such an oscillator we are going to introduce a canonical change of variables bringing 
the systems to the form of two decoupled linear oscillators.

For that, at first we integrate the first two equations of (\ref{e19}) representing the general solution in the 
following form
\begin{equation}
x=\alpha_1\tilde x_1+\tilde x_2, \quad y=\tilde y_1+\alpha_2\tilde y_2, 
\label{e21}
\end{equation} 
where
\begin{equation}
\tilde x_{1,2}=D_{1,2}\cos \Psi_{1,2} \quad \tilde y_{1,2}=D_{1,2}\sin \Psi_{1,2}, 
\label{e22}
\end{equation} 
$\Psi_{1,2}=\sigma_{1,2}t+\Psi^{0}_{1,2}$, $D_i$ and $\Psi^{0}_i$ are arbitrary constants, while eigenfrequencies
$\sigma_i$ can be found as solutions of a biquadratic equation
\begin{equation}
\sigma_{1,2}^{2}={1\over 2}(\omega_1^2+\omega_2^2+4\Omega^2\pm \sqrt{(\omega_1^2+\omega_2^2+4\Omega^2)^2-4\omega_1^2\omega_2^2}), 
\label{e23}
\end{equation} 
and
\begin{equation}
\alpha_i={2\Omega \sigma_i \over \omega_i^2-\sigma_i^2}, 
\label{e23a}
\end{equation}
where the indices $i=1,2$.

It is convenient to represent equations of motion in the horizontal direction in the canonical form 
introducing the corresponding Hamiltonian 
\begin{equation}
H={(P_1+\Omega y)^2\over 2}+{(P_2-\Omega x)^2\over 2}+{1\over 2}(\omega_1^2x^2+\omega_2^2y^2), 
\label{e24}
\end{equation}  
where $P_1$ and $P_2$ are canonical conjugates of $x$ and $y$, respectively.

Now one can prove by a direct substitution that when new coordinates, $\hat q_i$, and momenta 
$\hat P_i$, are introduced according 
to the rule
\begin{equation}
\tilde y_1={1\over f^{1/2}_1}\hat P_1, \quad \tilde x_1=-{\sigma_1\over f_1^{1/2}}\hat q_1, \quad \tilde x_2={1\over f_{2}^{1/2}}\hat P_{2}
\quad \tilde y_2={\sigma_2\over f^{1/2}_2}\hat q_2, 
\label{e25}
\end{equation} 
where 
\begin{equation}
f_i={\sigma_i^2(\sigma_i^2-\sigma_j^2)\over (\sigma_i^2-\omega_i^2)}, 
\label{e26}
\end{equation}
where $i\ne j$, the coordinate transformation (\ref{e21}) and (\ref{e25}) together with corresponding 
transformation of the momenta
\begin{equation}
P_1=-(\alpha_1\sigma_1+\Omega)\tilde y_1-(\sigma_2+\alpha_2\Omega)\tilde y_2, 
\end{equation}
\begin{equation} 
P_2=(\sigma_1+\alpha_1\Omega)\tilde x_1+(\Omega
+\alpha_2\sigma_2)\tilde x_2 
\label{e27}
\end{equation} 
provide a canonical transformation, which brings Hamiltonian (\ref{e24}) to the diagonal form
\begin{equation}
H=E_1+E_2, \quad E_i={1\over 2}({\hat P_i}^2+\sigma_i^2{\hat q_i}^2)={1\over 2}f_iD_i^2. 
\label{e28}
\end{equation}
Accordingly, the quantities
\begin{equation}
I_i={E_i\over \sigma_i}={1\over 2\sigma_i}f_iD_i^2 
\label{e29}
\end{equation}   
are the action variables. Therefore, they are adiabatic invariants.

\subsection{Equations for secular evolution of a star cluster  with a homogeneous stellar density}
\label{secular}

Following \cite{MitchellHeggie2007}, we
use the simple expressions for the gravitational potential discussed above.
It is assumed that initially, at a moment of time $t=t_0$, the cluster is situated far from galactic centre and the tidal effects as well as the ones due to the presence of Coriolis force can be neglected.  Also, we assume that initially the cluster has form of a sphere of radius $r_0$ and mass $M$, its initial density is $\rho_{0}={3M\over 4\pi r_0^3}$. Therefore, at $t=t_0$
we can set in equations (\ref{e18}) $\Omega=\gamma=0$. Due to the assumption of spherical symmetry
$A_1=A_2=A_3$ and it is easy to see from (\ref{e18}) that $\omega_{1}(t_0)= \omega_{2}(t_0)=\omega_{3}(t_0)\equiv \omega_0$, where 
\begin{equation}
    \omega_0=\sqrt{{4\pi\over 3}G\rho_0}.
\end{equation}
This quantity is used as a normalization factor in \S\ref{dimension}.

The orbit of the cluster assumed to be circular shrinks with time and, therefore, at later times
the tidal and Coriolis effect should be taken into account, the frequencies (\ref{e18}) are, in general, different from each other, the main
axes of ellipsoid, $a_1$, $a_2$ and $a_3$ are different from $r_0$ and the stellar density $\rho $ differs from $\rho_0$. 

It is the purpose of this Section to find out equations for the evolution of main axis and density
under the assumption of slowness of change of cluster orbit provided that initially their values 
are equal to $r_0$ and $\rho_0$, respectively. We consider the so-called $\beta$-model
of \cite{MitchellHeggie2007} where the amplitude $D_2$ defined in (\ref{e22}) is equal to zero for all stars retained
by the cluster. Additionally, it was shown in \cite{MitchellHeggie2007} that, for self-consistency,
the following relation
\begin{equation}
|\alpha_1|={a_1\over a_2},
\label{e30}
\end{equation}
where $\alpha_1$ is defined in (\ref{e23a}), should be satisfied for all times. This relation stems from the following arguments. The solution to the equation (\ref{e19})  describing vertical motion of a star 
can be written in the form $z=D_{3}\cos \Psi_3$, where $\Psi_3=\omega_3t+\Psi^{0}_3$, $D_3$ and $\Psi^{0}_3$
are constants of motion. Obviously, a star attains the maximal value of $z=D_3$ when $\cos \Psi_3=1$.
At these moments of time the orbit must touch the boundary of the ellipsoid, and, accordingly,
there should be ${x^2\over a_1^2}+{y^2\over a_2^2}+{z^2\over a_3^2}=1$. From (\ref{e21}) and (\ref{e22})
it follows that this condition can be rewritten in the form
\begin{equation}
{\alpha^2_1D_1^2\cos^2\Psi_1 \over a_1^2}+{D_1^2\sin^2 \Psi_1\over a_2^2}+{D_3^2\over a_3^2}=1.
\label{e31}
\end{equation}
Equation (\ref{e31}) must be satisfied for all values of $\Psi_1$, which is possible only 
when equation (\ref{e30}) is valid. In this case from (\ref{e31}) it follows that
\begin{equation}
{D_1^2\over a_2^2}+{D_3^2\over a_3^2}=1.
\label{e32}
\end{equation}
From eq. (\ref{e32}) it is seen that the maximal value of $D_1$ is $a_2$ and the maximal value 
of $D_3$ is $a_3$.

Now we express the adiabatic invariants $I_3$ and $I_1$ given by equations (\ref{e20}) and (\ref{e23}) through $a_2$ and $a_3$ assuming that the former invariant is evaluated for a trajectory with  the maximal $D_3$ and $D_1=0$, while the latter one is evaluated for a trajectory with the maximal $D_1$ and $D_3=0$. Since these quantities stay approximately constant during the evolution of our system 
they can be evaluated twice, for the initial moment of time and for some arbitrary time, thus 
linking values of the quantities of interest to the initial ones. We have
\begin{equation}
a_1=\sqrt{{2\omega_0 \sigma_1 \alpha_1^2\over f_1}}r_{0}, \quad a_2=\sqrt{{2\omega_0 \sigma_1 \over f_1}}r_{0}, \quad a_3=\sqrt{{\omega_0\over \omega_3}}r_0. 
\label{e33}
\end{equation}
Note the factor $2$ in the first and second expressions in (\ref{e33}). It appears because $f_1\rightarrow 2\omega_0$ in the limit
$\Omega \rightarrow 0$, see equation (\ref{e49}) below.  

Additionally, from the law of mass conservation we obtain the obvious relation
\begin{equation}
\rho ={\rho_0 r_0^3\over a_1 a_2 a_3}. 
\label{e34}
\end{equation}
Equations (\ref{e33}) and (\ref{e34}) are the evolution equations of our model. In general,
they must be solved numerically, since values of main axes enter r.h.s implicitly, through
the dependency of the coefficients $A_1$, $A_2$ and $A_3$ on them. Note that the solutions
should be different from the solutions of an analogous incompressible model. This 
difference stems from the fact that the analogue of pressure, velocity tensor $<v_iv_j>$,
where brackets stand for averaging with a distribution function in phase space, is not 
zero at the surface in the stellar dynamical model.     
 





\section{Solutions of the secular equations}
Based on the above formalism, we examine the condition under which the tidal perturbation from the galactic potential is 
compressive.  Around a point mass potential, a cluster would be tidally disrupted if its galactic orbital frequency $\omega$
is larger than its characteristic internal frequency $\omega_0$.  However, around a galaxy with a shallow density distribution,
a cluster may preserve its integrity deep in the galactic potential where $\omega > > \omega_0$.  We first consider an idealized
case of negligible $\gamma^2$ which corresponds to a homogeneous density distribution of galactic stars. 
We show that the compression by the galactic tide preserve the spherical shape of the 
cluster.  When first order contribution of a small 
$\gamma^2$ is taken into account, we identify the condition for tidal disruption in terms of ratio between 
$\gamma^2$ and angular frequency $\Omega^2$ (\S3.2).  We introduce a idealized 
power-law density for the galaxy and estimate the critical radius $r_t$ outside which a cluster would withstand tidal 
disruption (\S3.3).  In the limit of small $\gamma^2$, we show that the tidal perturbation from a background potential
due to a relatively flat density distribution is predominantly compressive (\S3.4). For the Milky Way, we suggest the disk
contribution to the tidal field, if dominant, can ensure the survival of migratory stellar cluster (\S3.5).

\subsection{Natural units}
\label{dimension}

In what follows it it convenient to express all quantities of the dimension of a frequency 
entering the problem apart from $\gamma $, such as $\sigma_i$, $\omega_i$, $\Omega$ in units of $\omega_0$, semimajor axes $a_i$ in units of $r_0$ and density in units of $\rho_0$. This will be implicitly implied hereafter.

We also introduce the ratio of $\gamma$ to angular frequencies of the cluster's orbit around the galaxy,
\begin{equation}
\tilde \gamma \equiv \gamma /\Omega. 
\label{e37}
\end{equation}
For a point mass galactic potential, it is 
$\sqrt{3}$.  But, we are considering potentials for galaxies with relatively shallow density
distribution.  In this case, $\tilde \gamma $ can be treated as a small parameter and a simple analytic solution of the secular 
equations is possible.

\subsection{The limit of a strongly compressed star cluster} 
 
At first let us consider a star cluster situated deep within the potential well of a galaxy assuming that
$\Omega \gg 1$. Note that the condition $\Omega > 1$ may be used as the tidal disruption condition in the standard situation
when $\tilde \gamma \sim 1$. We assume, however, that $\tilde \gamma $ is small and may be neglected in the leading approximation. In this limit equation (\ref{e23}) tells that $\sigma_1 \approx 2\Omega $ and $\sigma_2\approx 0$. In this case it is seen from equations (\ref{e23a}) and (\ref{e26}) that we have $\alpha_1 \approx -1$ and $f_1 \approx 4\Omega^2$. 
Using equation (\ref{e30}) we find that $a_1 \approx a_2$, while equations (\ref{e33}) tell that $a_1 \approx a_2 \approx a_3 \approx {1\over \sqrt{ \Omega }}$.  In summary, in the leading approximation a strongly compressed star cluster maintains its spherical form with both $\phi $ 
and $\theta $ being small and 
\begin{equation}
a \equiv  a_i \approx {\Omega}^{-1/2}, \quad \rho \approx  {\Omega}^{3/2}. 
\label{e38}
\end{equation}     

The next order corrections taking into account effects of non-zero $\tilde \gamma $ and self-gravity can be easily found using the fact that, for a spherical cluster, all $A_i$ in equation (\ref{e18}) are equal to $2/3$, and that we can can use these and the expressions (\ref{e38}) when considering $\omega_i$ in equation (\ref{e23}), 
since these characteristic frequencies are assumed to be much smaller than $\Omega$. We obtain 
from (\ref{e18}) $\sigma_{1} \approx 2\Omega (1+{{\omega_1}^2+{\omega_2}^2\over 8})$. Equations (\ref{e30}) and
(\ref{e33}) can be used again to find the corrected values of $a_i$ and $\rho$. Since calculations are straightforward
we show only the result:
\begin{equation}
a_1={ 1 \over {\Omega}^{1/2}} \left(1-{1\over 4\Omega^{1/2}}+{{\tilde \gamma}^2\over 16}\right), 
\end{equation}
\begin{equation}  
a_2={1 \over \Omega^{1/2}}\left(1-{1\over 4\Omega^{1/2}}+{3{\tilde \gamma}^2\over 16}\right), 
\end{equation}
\begin{equation}  
a_3={ 1 \over {\Omega}^{1/2}} \left(1-{1\over 4{ \Omega}^{1/2}}\right), 
\end{equation}
\begin{equation}  
\rho \approx  {\Omega}^{3/2}
\left(1+{3\over 4 \Omega^{1/2}}-{{\tilde \gamma}^2\over 4}\right). 
\label{e39}
\end{equation}   
From the above Equations, it is seen that the corrections are small when $\tilde \gamma < 1$ and $\Omega \gg 1$, and that $a_1 < a_2$,
i. e. the cluster elongation in the direction of motion is larger than the one in the direction of the galactic centre. 
This orientation is orthogonal to that of the analogous incompressible fluid model, where the axis is elongate in the direction
of the galactic center.

Although the corrections get smaller with an increase of $\Omega $ when it gets sufficiently large $\omega_1$ defined in equation (\ref{e18}) 
becomes imaginary, which results in runaway of stars from the cluster and its disruption. Equating $\omega_1$ to zero and using $A_1=2/3$ and
the expressions (\ref{e38}) we find a very simple criterion of tidal disruption of a cluster in our model - the cluster is disrupted when
\begin{equation}
\tilde \gamma \equiv \gamma/\Omega > \Omega^{-1/4}.
\label{e40}
\end{equation}        

\subsection{A simple model of galactic tidal field}
\begin{figure}[ht!]
\plotone{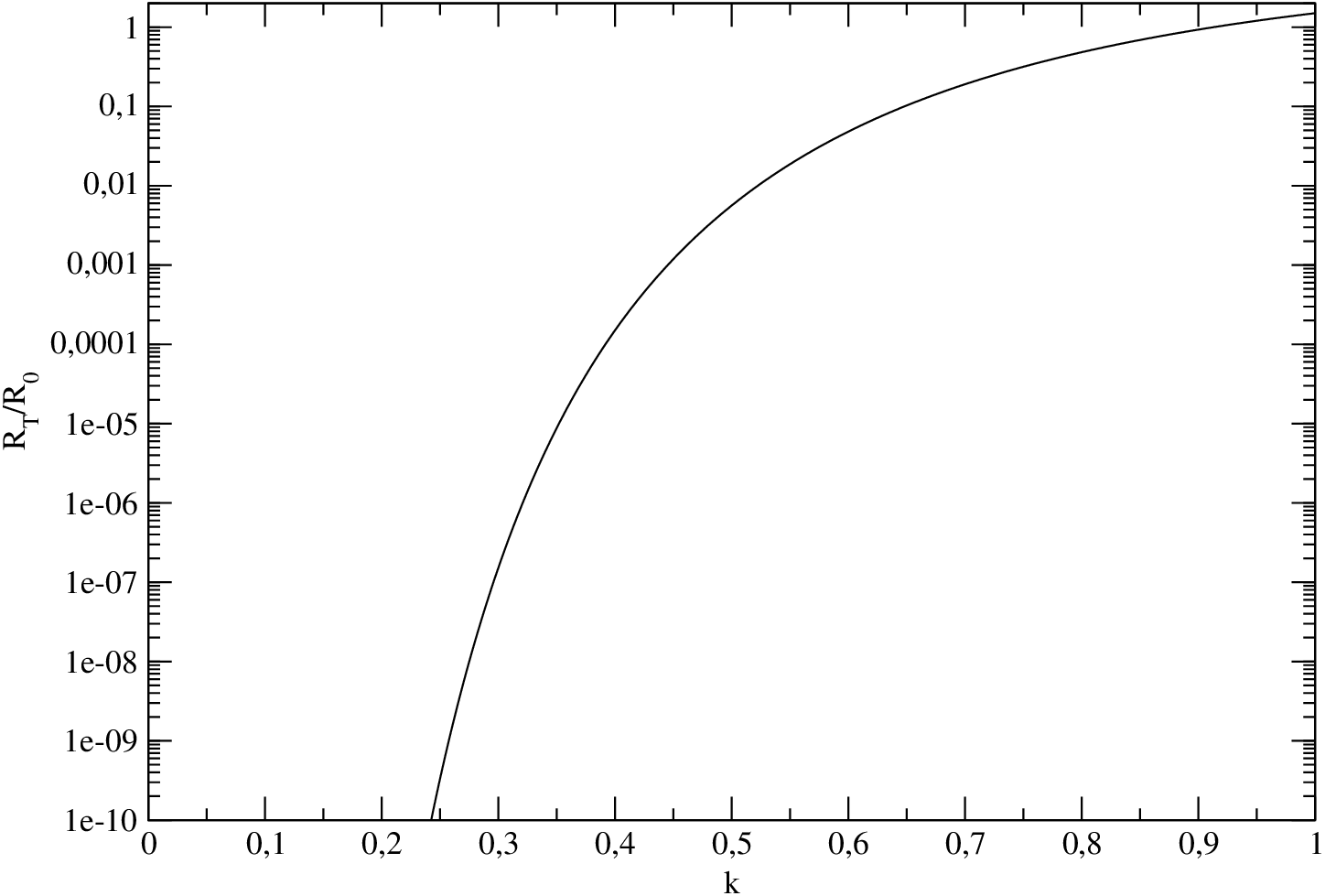}
\caption{We show the dependence of $R_t$ on $k$ given by equation (\ref{e43}).}\label{fig1}
\end{figure}

As an example of distribution of galactic density let us consider a power law dependence
\begin{equation}
\rho_G=\rho_0(R/R_0)^{-k},
\label{e41}
\end{equation} 
noting that it is normally expected that a cluster would be disrupted at $R\sim R_0$. From equations (\ref{e3}) it follows that in case of distribution
(\ref{e41}) we have
\begin{equation}
\tilde \gamma = k^{1/2}, \quad \Omega = \left({3\over 3-k}\right)^{1/2}{\tilde R}^{-k/2},
\label{e42}
\end{equation}
where $\tilde R = R/R_0$. From our criterion for tidal disruption (\ref{e40}) it follows that the cluster is disrupted when $R < R_T$, where
\begin{equation}
R_{T}={\left({3\over 3-k}\right)}^{1/k}k^{4/k}R_{0}.
\label{e43}
\end{equation}
It is very interesting to note that according to the criterion (\ref{e43}) the cluster is practically indestructible even when $k$ is not very
small (see Fig. \ref{fig1}). Say, when $k=0.5$ we have $R_T\approx 6\cdot 10^{-3}R_0$.

\bigskip
\bigskip    
\bigskip
\bigskip
\bigskip
\bigskip
\bigskip
\bigskip
\bigskip
\bigskip
\bigskip
\bigskip

\subsection{An analytic solution of the secular equations in the limit of small $\tilde \gamma$}

Equation (\ref{e39}) tells that when $\tilde \gamma =0$ cluster remains spherical in the limit of strong compression $\Omega \gg 1$. On the other hand, it is obviously spherical when the tidal field is absent and $\Omega=0$. This suggest that it is reasonable to assume that it is spherical
when $\gamma=0$ for any value of $\Omega$. We are going to show that this is indeed the case and consider the following ansatz for the axis 
$a_i$ and the frequencies $\omega_i$
\begin{equation}
a_i=a_0(1-\delta_i), \quad \omega^2_{1,2}=a_0^{-3}+\Delta_{1,2}, \quad \omega^2_3=a_0^{-3}+\Delta_3+\Omega^2,
\label{e44}
\end{equation}  
where it is implied that both $\delta_i$ and $\Delta_i$ are small being proportional to $\tilde \gamma^2$. Substituting the expressions 
for the frequencies into (\ref{e18}) and (\ref{en6}) and taking into account (\ref{e34}) we get
\begin{equation}
\Delta_{1}={3\over 5a_0^3}(\delta_2+\delta_2+3\delta_1)-\gamma^2, \quad \Delta_{2,3}={3\over 5a_0^3}(\delta_{1,2}+\delta_{3,1}+3\delta_{2,3}).
\label{e45}
\end{equation}

Now we substitute (\ref{e45}) in (\ref{e23}) to obtain
\begin{equation}
\sigma^{2}_{1,2}=(\omega_{*}\pm\Omega)^{2}
\left(1+{\Delta_1+\Delta_2\over 4\omega_*(\omega_*\pm \Omega)}\right),
\label{e46}
\end{equation}
where
\begin{equation}
\omega_{*}=\sqrt{\Omega^{2}+a_0^{-3}}.
\label{e47}
\end{equation}
Note that when $\tilde \gamma =0$ $\omega_*=\omega_3$.

We substitute (\ref{e45}) in (\ref{e23a}) and (\ref{e26}). From (\ref{e23a}) we get
\begin{equation}
\alpha_1^2=1+{\Delta_1-\Delta_2\over 2\Omega(\omega_*+\Omega)},
\label{e48}
\end{equation}
and from (\ref{e26}) we get
\begin{equation}
{2\sigma_1\over f_{1}}={1\over \omega_*}
\left(1+{\Delta_2-\Delta_1\over 4\Omega (\omega_*+\Omega)}-{\Delta_1+\Delta_2\over 4\omega_*^2}\right).
\label{e49}
\end{equation}
Note that in the limit $\Omega \rightarrow 0$ we have $\sigma_1, \omega_* \rightarrow \omega_0$, and, therefore, $f_1\rightarrow 2\omega_0$.
This explains the factor $2$ in (\ref{e33}). From the expression for $\omega_3$ and the definition of $\Delta_3$ 
we get $\omega_3=\omega_*(1+{\Delta_3\over 2\omega^2_*})$.

\begin{figure}[ht!]
\plotone{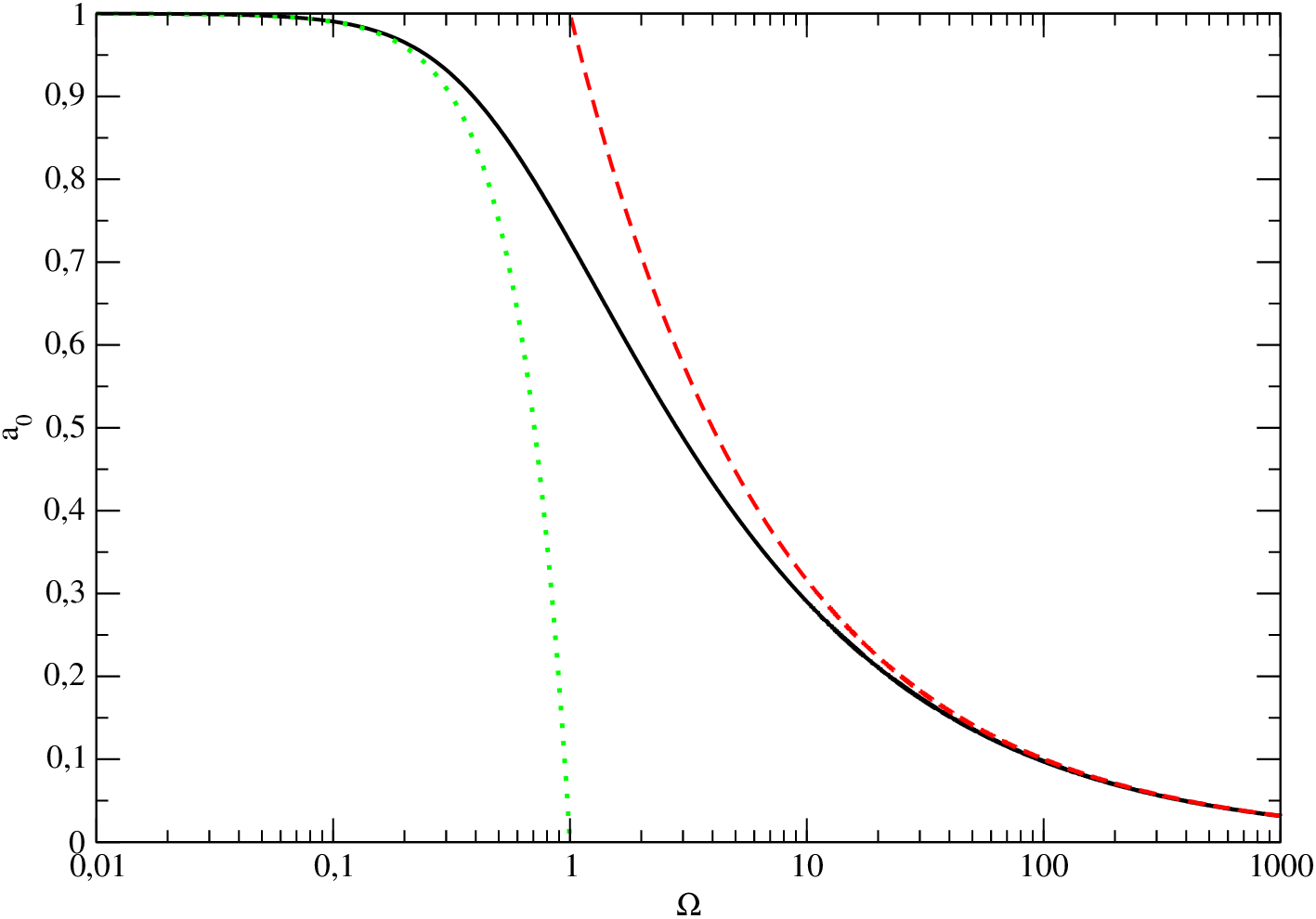}
\caption{The result of solution of equation (\ref{e50}) together with the corresponding approximate expressions. See the text for
a description of different curves.}\label{fig2}
\end{figure}

Now we substitute the expressions above into the secular equations (\ref{e33}). All equations (\ref{e33}) result in only one 
zero order equation for the quantity $a_0$
\begin{equation}
a_0^{2}={1\over \omega_*}={1\over \sqrt {\Omega^2 + a_0^{-3}}}.
\label{e50}
\end{equation}
This justifies our assumption that when $\tilde \gamma =0$ cluster remains spherical for all values of $\Omega$. It is obvious 
that (\ref{e50}) results in a quartic equation for $a_0$ with coefficients depending only on $\Omega$. The physically acceptable  
solution of this equation is shown in Fig. \ref{fig2} as a solid line. As a dashed line we show the corresponding asymptotic solution in the limit of large $\Omega$, $a_0\approx \Omega^{-1/2}$ and as a dotted line the approximate solution in the limit of small $\Omega$,
$a_0\approx 1-\Omega^2$ is shown.

\bigskip
\bigskip
\bigskip
\bigskip
\bigskip
\bigskip
\bigskip
\bigskip
\bigskip
\bigskip
\bigskip
\bigskip
\bigskip
\bigskip
\begin{figure}[ht!]
\plotone{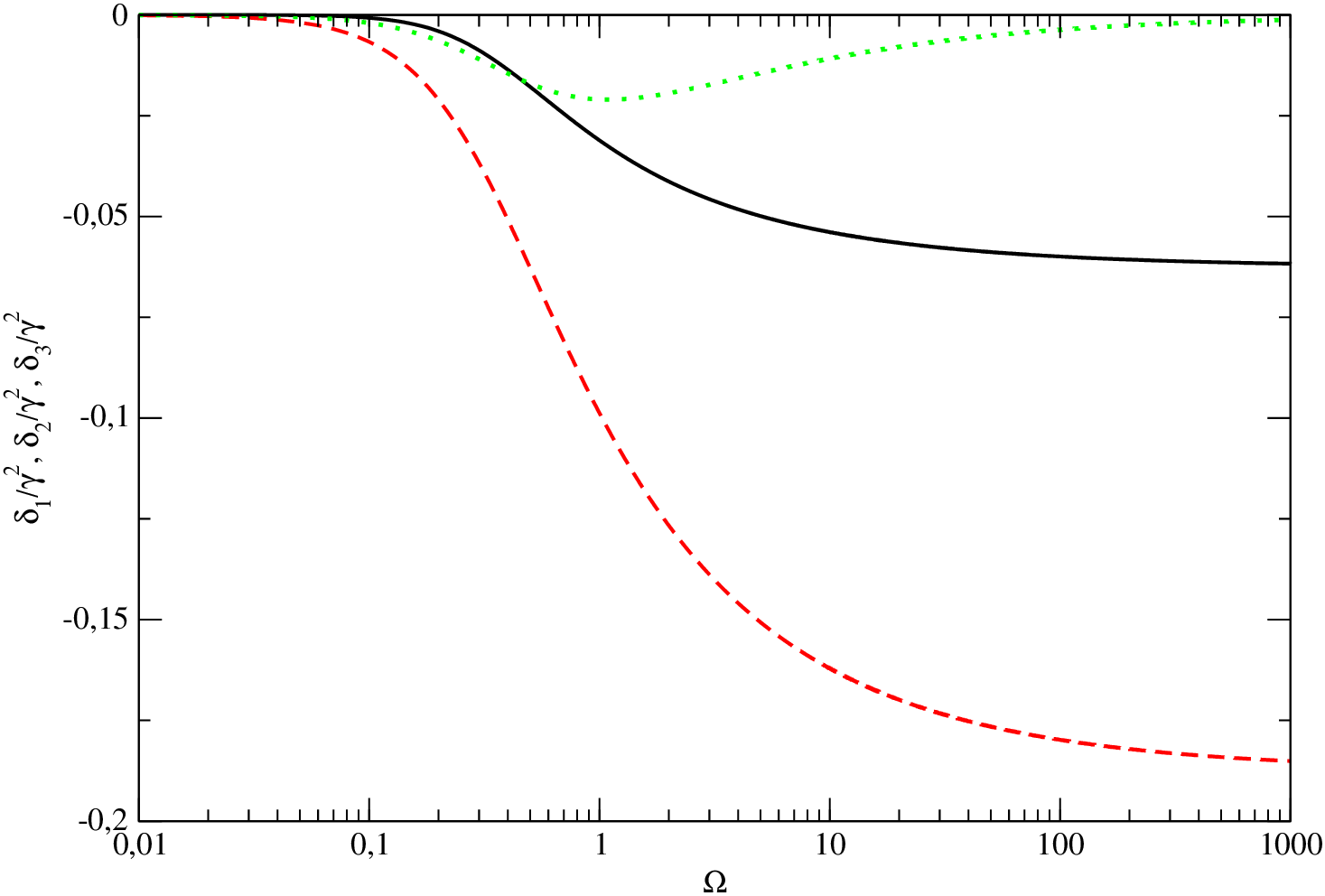}
\caption{We show solutions (\ref{e52}), (\ref{e53}) of equations (\ref{e51}). See the text for a description of different curves.        }\label{fig3}
\end{figure}
\bigskip
\bigskip
\bigskip
\bigskip
\bigskip
\bigskip

The perturbed part of the secular equations can be compactly written introducing new variables $\delta_{\pm}=\delta_1\pm \delta_2$ and
 $\Delta_{\pm}=\Delta_1\pm \Delta_2$. We have
\begin{equation}
\delta_{3,+}={\Delta_{3,+}\over 4\omega_*^2}, \quad \delta_{-}={\Delta_{-}\over 4\Omega (\omega_*+\Omega)}.
\label{e51}
\end{equation}
We substitute (\ref{e45}) into (\ref{e51}) to obtain
\begin{equation}
\delta_{-}={\delta_*\over {\Omega^2a_0^4+\Omega a_0^2+{3\over 10}a_0}}, 
\quad \delta_{+} = -{(1+{9\over 10}a_0)\delta_* \over 1-{21\over 20}a_0+{9\over 40}a_0^2}, 
\label{e52}
\end{equation}
\begin{equation} 
\delta_3=-{3 \delta_*\over 20(1-{21\over 20}a_0+{9\over 40}a_0^2)}, \ \ \ \ {\rm where} \ \ \ \ 
\delta_*={\tilde \gamma^2\Omega^2\over 4\omega_*^2}.
\label{e53}
\end{equation}
The original quantities can be easily recovered from the obvious relations $\delta_{1,2}={1\over 2}(\delta_{+}\pm \delta_{-})$. 
When considering the limit $\Omega \rightarrow \infty$ it is possible to show that the expressions (\ref{e52}) and (\ref{e53}) 
give corrections proportional to $\tilde \gamma^{2}$, which are in agreement with the previous result (\ref{e39}). 

It is seen from (\ref{e50}), (\ref{e52}) and (\ref{e53}) that the ratios $\delta_i/\tilde \gamma^2$ are the functions of 
$\Omega$ only. We represent them in Fig. \ref{fig3} for $\delta_1$, $\delta_2$ and $\delta_3$ shown as solid, dashed and 
dotted lines, respectively. One can see from this Fig. that all $\delta_i$ are negative. Thus, the presence of 
a non-zero, but small $\gamma $ always leads to a small expansion of the cluster as expected. It is also seen that
the absolute value of $\delta_3$ is larger than that of $\delta_2$ when $\Omega \le \Omega_*\approx 0.46$. When $\Omega \rightarrow
\infty$ $\delta_3\rightarrow 0$.  

\section{Applications to some empirical galaxy models}


\subsection{Parametric model potentials}


In the analysis of observational data for the central regions of galaxies, a frequently used prescription is the modified
Hubble profile \citep{Cote2006} in which the density at $R<2 R_K$ can be approximated as 
\begin{equation}
    \rho_G= \rho_K (R) \simeq {\rho_c \over (1 + R^2/R_K^2)^{3/2}}
\end{equation}
where $R_K =\sqrt {9 \sigma^2 / 4 \pi G \rho_c}$ is the King radius, $\rho_c$ and $\sigma$ are the central density 
and velocity dispersion\citep{Binney2008}.  Since $\rho_K$ is approximately homogeneous and $k \sim 0$, the tidal 
perturbation in compressive.  But in the outer regions of the King model, $\rho_K (R) \propto R^{-3}$ and the 
tidal perturbation is disruptive.

For bulge of disk galaxies and elliptical galaxies, the classical \citet{Jaffe1983} potential is generated
from a density distribution
\begin{equation}
    \rho_G = \rho_J (R) = {\rho_B \over (R/R_B)^2 [1+(R/R_B)^2]}
\end{equation}
where $\rho_B$ is a normalized density and $R_B$ is the scaling parameter.  For $R < < R_B$ $\rho_J (R) \propto R^{-2}$
and $k=2$ so that the tidal perturbation is disruptive.

Another frequently used \citet{Hernquist1990} potential is generated from a density distribution
\begin{equation}
\rho_G= \rho_H (R) = {\rho_{H0} \over (R/R_H) [1 + (R/R_H)]^3 }
\end{equation}
where $\rho_{H0}$ is a normalized density and $R_H$ is the scaling parameter.  For $R < < R_H$ $\rho_H (R) \propto R^{-1}$
and $k=1$ so that the tidal perturbation is disruptive.

A more general $\eta$ potential \citep{Tremaine1994}, the associated density distribution is
\begin{equation}
\rho_G=\rho_\eta (R) = {\eta \rho_{\eta 0} \over (R/R_\eta)^{3-\eta} [1 + (R/R_\eta)]^{1 + \eta} }, \ \ \ \ \ \ \ \ 0 < \eta \leq 3
\end{equation}
where $\rho_{\eta 0}$ is a normalized density, $R_\eta$ is the scaling parameter, and $\eta$ is a power index parameter.  
For $R < < R_\eta$, $\rho_\eta (R) \propto R^{\eta-3}$ and $k=\eta-3$ such that it is reduces to the King, Herquist,
and Jaffe model with $\eta=3, 2,$ and 1 respectively.  Moreover, contribution from the point-mass potential of
SMBH can be added to the $\eta$ potential.  Depending on the value of $\eta$, the tidal perturbation can 
be both compressive and disruptive.

\subsection{Empirical S{\'e}rsic models}

The surface brightness $I$ of elliptical galaxies and the bulge of spiral galaxies is commonly modeled
\citep{Kormendy2009} in terms of an empirical \citet{Sersic1968} profile with $I=I(0) \mathrm{exp} 
[- b_n (D/R_S)^{1/n}]$ where $I(0)$ is central 
surface brightness, $D$ is the projected distance from the center, $R_S$ is scaling radius, $b_n=2n-0.324$, and 
$1 \leq n \leq 15$ is the fitting power index.  For a spheroid, the associated density at a distance $r$ from the 
galactic center can be approximated \citep{Prugniel1997, Terzic2005} by
\begin{equation}
    \rho_G= \rho_S (R) = \rho_{S0} (R/R_S)^{-p_n} {\mathrm{exp}} [-b_n (R/R_S)^{1/n}]
\end{equation}
where $\rho_{S0}$ is a normalization constant.  The power index can be approximated as
$p_n=1-0.6097/n+ 0.05563/n^2$ for $0.6 \leq n \leq 10$ and $10^{-2} \leq R/R_S \leq 10^3$.
Observational fit\citep{Graham2005} show that the magnitude of $n$ increases from $0.5$ 
to $10$ for galaxies mass in the range of $10^7 - 10^{12} M_\odot$.  At the low mass end
$k \sim 0$ and the tidal perturbation is compressive whereas for the massive elliptical
galaxies (with $n$ approaching 10), $\rho_S(R) \propto R^{-1}$ near the center so that
the tidal perturbation is disruptive.  
We thank an anonymous referee for putting out to us the
trend of Sersic index most likely applies to the outer slopes, not the inner regions 
and more massive galaxies may have cores with lower central densities.
This correlation may account for the dichotomy
between the presence of nuclear clusters around galaxies with comparable or less mass
than the Galaxy and their absence in massive elliptical galaxies.  

\subsection{Galactic potential}
There are several empirical prescriptions for the gravitational potential of the Galaxy.  In general, contribution to 
$\Phi_G$ is considered to be the sum of that due to the central SMBH ($\Phi_{SMBH}$),  the Galactic bulge ($\Phi_{bulge}$),
the Galactic disk ($\Phi_{disk}$), and the halo ($\Phi_{halo}$) \citep{Gnedin2005, Widrow2005} where
\begin{equation} 
    \Phi_G=\Phi_{SMBH}+\Phi_{bulge}+\Phi_{disk}+\Phi_{halo}, \\
    \end{equation}
    \begin{equation} 
    \Phi_{SMBH}=-GM_{SMBH}/R, \\
    \end{equation}
    \begin{equation} 
    \Phi_{bulge} = -G M_{bulge} /(R+R_{bulge}), \\
    \end{equation}
    \begin{equation} 
    \Phi_{disk} = -G M_{disk}/ [(\sqrt{(z^2 + b^2)}+a)^2+\varpi^2]^{1/2}, \\
    \end{equation}
    \begin{equation} 
    \Phi_{halo} = -G M_{halo} {\rm ln} (1 + R/R_{halo}).
\end{equation}
where $R$, $\varpi$, and $z$ is the total distance, in the disk radius, and distance above the disk;
$R_{bulge} (=0.6$ kpc), $a (=5$ kpc), $b (=0.3 $ kpc), and $R_{halo} (=20 $ kpc) are the scaling 
length for the bulge, disk, and halo respectively; $M_{SMBH} =4 \times 10^6 M_{\odot}$ is the 
mass of the SMBH, $M_{bulge} (=10^{10} M_\odot)$, $M_{disk} (=4 \times 10^{10} M_\odot)$, 
and $M_{halo} (=10^{12} M_\odot)$ are the mass scaling
factor for the bulge, disk, and halo respectively\citep{Miyamoto1975, Hernquist1990,  Navarro1997,  Dehnen1998, YuMadau2007}.
Various values of these model parameters are summarized in \citet{Kenyon2008}.

From the Poisson equation, we find the corresponding density which contributes to these components of
the potential:
\begin{equation} 
    \rho_{bulge}= {M_{bulge} \over 4 \pi R_{bulge}^3} {(R_{bulge}/R -1) \over (R/R_{bulge}+1)^3}, \\
    \end{equation}
    \begin{equation} 
    \rho_{disk} = {M_{disk}  \over 4 \pi [\varpi^2 + (a+\sqrt{b^2 + z^2})^2]^{3/2}}   \\
    \end{equation}
    \begin{equation}{*} 
    \left[ {a \over \sqrt{b^2 +z^2}} +{3 (a +\sqrt{b^2 +z^2})^2 \over
    \varpi^2 + (a+\sqrt{b^2 +z^2})^2}-{3 z^2
    (1 + {a / \sqrt{b^2+z^2}})^2 \over \varpi^2 + (a+\sqrt{b^2 +z^2})^2} \right], 
    \end{equation}{*}
    \begin{equation} 
    \rho_{halo} = {M_{halo} \over 4 \pi R^3} \left[ {R (2 R+R_{halo}) \over (R+R_{halo})^2} - {\rm ln} \left(1+{R\over R_{halo}}\right) \right].
    \end{equation}
Deep in the galactic potential where $R$, $\varpi$, and $z$ are relatively small compared with other scaling
parameters, $\rho_{bulge} \propto R^{-1}$, $\rho_{disk} \sim$ constant, $\rho_{halo} \propto R^{-2}$.
Only the density associated with the disk potential become slowly varying functions of $R$ and 
$z$ with $k < < 1$, $\gamma < < \Omega$, and compressive tidal perturbation. This contribution is
negligible over most regions of the Galaxy including the proximity of  \sgra.  In most regions of 
the present-day Galaxy, the dominant tidal perturbation from other components (SMBH, bulge, and halo)
are disruptive.  Nevertheless, during the galactic infancy, after the formation of the disk and prior 
to the formation of a substantial bulge or central black hole, it is possible for stellar clusters
to retain their integrity on their migratory routes to the galactic center. 

\subsection{An estimate of inspiral 
time scale in case of galactic centres with shallow
density profiles}

Nuclear clusters arrive in the galactic center under the action of dynamical friction.   
In this subsection, we estimate the clusters' typical in-spiral timescale $T_{DF}$.
For the galactic background potential, we use a general power-law density distribution 
(\ref{e41}), which, as follows from the previous section can be used to describe many expected 
density profiles in inner parts of galaxies. To find $T_{DF}$ we use the expression (8.9) 
in \citet{Binney2008} to determine the absolute value of force appearing due to the effect
of dynamical friction, $\tilde F$.  Whereas the original calculation is appropriate for the 
case of $k=2$, we modify $\tilde F$ for a generalized power-law density distribution such that
\begin{equation}
 \tilde F\approx 5.38 \ln \Lambda {G^2 M^2 \rho_G\over V^2},
 \label{en1}   
\end{equation}
where, to be consistent with the notation in other sections, $M$ is the cluster mass, 
$\ln \Lambda $ is the Column logarithm and $V=\Omega R$ is the cluster's orbital velocity. 
Note that the orbit is assumed to be nearly circular 
during the whole orbital evolution. This assumption 
may not actually be valid for the shallow density profiles,
since in this case the orbital eccentricity may grow 
\citep{Polnarev1994, Vecchio1994}.  However, we neglect
this effect here assuming that it wouldn't significantly change our
order of magnitude estimates. 

In the above expressions, the representative
density for the cluster ($\rho$) is formally less than
that of the galactic background ($\rho_G$) at location $R 
< R_0$ (where they are equal).  However, the total stellar
density within the volume occupied by the clusters' stars
is the sum of bound cluster stars and that of the galactic
stars which merely pass through the cluster. Physically,  
$\rho$ represents an {\it overdensity}, and, accordingly,
mass $M$ is the mass excess.

In order to estimate the dynamical friction timescale  
\begin{equation}
T_{DF}    \sim {MV\over \tilde F},
\label{en2}
\end{equation}
we need to specify the mass and spatial scales for both 
the cluster and its host galaxy.  For galaxies with
density profile (\ref{e41}), we scale $\rho_0$ 
in terms of a reference mass $M_G$ at a given radius $R_\ast$.  
For galaxies similar to the Milky Way\citep{Kenyon2008}, 
we scale $M_G$ and $R_\ast$ by dimensionless parameters 
$M_9=M_G/10^9M_{\odot}$ and $R_2=R_\ast/10^2$pc. 
Typical values of $M_9$ and $R_2$ are of the order unity.
We also scale the cluster's mass $M$ and radius $r_0$ by 
dimensionless parameters $m_5=M/10^5$ and $r_1=r_0/10$ pc 
respectively. In the scaling in physical units, $r_0$ 
corresponds to the half mass radius of realistic clusters.
Using the above notations, the density profile
(\ref{e41}) can be represented as
\begin{equation}
\rho_G (R) = 10^3{(3-k)\over 4\pi}
\left({M_9\over R_2^3}\right)
\left({R\over R_*}\right)^{-k}{M_{\odot}\over pc^3}
\label{en3}    
\end{equation}
and the typical internal dynamical timescale associated with the
cluster, 
\begin{equation}
    \omega_0^{-1}=\sqrt{{R_0^3\over G M}}\approx 1.4
    \times 10^6 \left({r_1^3\over m_5} \right)^{1/2} \ {\rm yr}.
\end{equation}
 
We substitute (\ref{en1}) and (\ref{en3}) in (\ref{en2})
and take into account (\ref{e42}) to obtain
\begin{equation}
T_{DF} (R) \approx {1.75\times 10^8 \over \Lambda_{20}}
{3 M_9 \over (3-k)m_5}
\left({R_2^3\over M_9}\right)^{1/2}\left({R\over R_*}\right)^{3-k/2} \ {\rm yr},
\label{en4}
\end{equation}
where we assume that a typical value of $\Lambda$ is
order of $\ln(10^9)\sim 20$ and $\Lambda_{20}=\Lambda/20$.
Equation (\ref{en4}) tells that the dynamical friction time
is reasonably fast for the considered values of numerical
parameters, but it sharply grows with $R$.

It is of interest to compare our tidal disruption 
radius $R_T$ given by (\ref{e43}) with $R_*$. Since
the condition $\rho_G(R)=\rho_0$ defines the characteristic radius $R_0$, we find
\begin{equation}
R_0=\left[10 \left({3-k\over 3}\right)\left({M_9\over m_5}\right)\left({r_1^3\over R_2^3}\right) \right]^{{1\over k}}R_*.
\label{en5}    
\end{equation}
Substituting (\ref{en5}) in (\ref{e43}),  we have
\begin{equation}
 R_T=\left[10k^4 \left({M_9\over m_5}\right)\left({r_1^3\over R_2^3}\right)\right]^{{1\over k}}R_*.
 \label{enn6}   
\end{equation}
Thus, the condition $R_T < R_*$ results in
\begin{equation}
k < 0.1^{1\over 4}\left({m_5R_2^3\over M_9r_1^3}\right)^{{1\over 4}}
\approx 0.56 \left({m_5R_2^3\over M_9r_1^3}\right)^{{1\over 4}}.
\label{en7}    
\end{equation}
This condition rather weakly depends on the ratio of typical densities of the cluster and galaxy.
For the considered model parameters, it is typically satisfied. For example, if we adopt
the nominal values of these scales with $k=0.5$, we would find $R_T\approx 0.4R_* \approx 40pc$.

\section{Summary and Discussions}

It has long been assumed that tidal perturbation on satellites (including individual stars or
stellar clusters) by an external gravitational field is disruptive.  However, the conventional 
tidal disruption radius is derived for a point mass back ground potential.  This approximation
may not be appropriate for a general mass distribution.  

In this paper, we examine the tidal stability of stellar clusters in a background gravitational 
potential with a power-law density distribution.  In order to gain some physical insight, we
construct an analytic formalism with some idealized assumptions.  We consider a cluster with a
homogeneous internal density and a circular orbit around an spherically symmetric slowly 
varying background galactic potential.  This approximation is analogous to the classical 
theory of uniform ellipsoidal figures under tidal perturbation of a companion \citep{Chandrasekhar1969}.
The advantage of this approach is that it enabled us to analytically obtain the shape of the cluster
and the stellar orbits inside it.  We also use these analytic solution to identify adiabatic 
invariants which can be used to extrapolate the cluster's adiabatic response from negligible 
to strong tidal field through slow (compared with the cluster's internal dynamical time scale) evolution.
Similar approach has been used by \citep{Young1980}
in his consideration of the  adiabatic black hole growth, a similar problem was also recently considered in \citet{Jingade2016} for
 S{\'e}rsic Models of Elliptical Galaxies.

With this method, we calculate the condition under which the stellar orbits become unstable.  
We show that if the galactic density distribution is a weakly decreasing power-law function of
radius, the cluster can preserve its integrity at radii much smaller than the conventional tidal
radius, i.e. the cluster can survive deep in the gravitational potential of the galaxy.  Although
the density inside the survivable clusters is comparable to that of the galactic background, we 
suggest their accumulation can lead to the gradual build up of the nuclear clusters.

There are several potential observational tests.  The effect of tidal compression
enables the clusters to retain their internal velocity dispersion as they undergo orbital decay
towards the center of their host galaxies.
1) The nuclear clusters formed along this channel are likely to preserve their velocity dispersion 
and it is generally smaller than that of
the surrounding field stars \citep{geha2002}.  2) When multiple clusters reach the galactic nuclei, the peak of 
their composite 
surface density may be slightly displaced from the galactic center.  Both of these two dynamical
effects have already been suggested and shown through some preliminary simulations 
by \citet{Oh2000b}.  3) If the progenitors of the nuclear clusters originated 
from the galactic halo with sub-solar heavy element abundance, similar to that of the Galactic 
globular clusters, their convergence at the galactic centers would enhance the nuclear 
clusters' metallicity dispersion in contrast to that of the surrounding field stars.
However, old and metal-deficient stars transported by the preserved stellar clusters may be outshine
by the recently-formed young and metal-rich stars, especially in active galactic nuclei \citep{Artymowicz1993}.

Our results also show that if background density falls off faster than $r^{-1}$, the classical
tidal radius may still apply.  This disruptive effect would occur if the tidal field is dominated
by the point mass potential of super massive black holes or possibly by that of galactic bulges.
We speculate this dichotomy may be the cause of 1) mutual exclusion between nuclear clusters and
supermassive black holes in the center of massive elliptical galaxies and 2) the dominance of 
nuclear clusters over black holes in galaxies where they coexist, as in the case of the Milky way.

Our analytic approach is particularly useful to highlight the basic physical effects.
Nevertheless, it is based on idealized 
models of stellar clusters and adiabatic extrapolation.
These models may suffer from many potential instabilities. It is not clear \footnote{For the similar 
Freeman bar models such instabilities were considered in \citep{Morozov1974} and \citep{Tremaine1976c}} whether 
these instabilities are physically generic or reflect the very simplified nature of our approach. 
Although it is technically challenging to extend the analytic analysis to more realistic models with a similar approach. 
there is a simple argument, which enables us to postulate that such models may behave qualitatively in the similar way. 
Namely, when $\tilde \gamma$ is small and the cluster is deep within the potential well of a galaxy so that $\Omega \gg 1$,
the cluster's dynamics should be determined by only this frequency. In particular, an orbital period of a 'typical' star 
should be order of $\Omega^{-1}$ and, accordingly, its energy (per unit of mass) and the corresponding 'typical' action 
should be $E \sim a^2\Omega^2$ and $I\sim a^2\Omega$, respectively, where $a$ is a characteristic size of the cluster. 
Since the action is conserved we have $a \sim \Omega^{-1/2}$. Now, from equation (\ref{e1}) it follows that the condition 
that the combination of tidal and centrifugal forces in the $x$ direction exceeds
self-gravity force can be approximately formulated as $\tilde \gamma^2 \Omega^2  > {Gm \over a^3}$, where we temporarily
restore the physical units. Going back to the natural units
and substituting $a=\Omega^{-1/2}$ in the condition we have
again our tidal disruption criterion (\ref{e40}). Note that,
perhaps, this argument can be obtained in a more rigorous way using the formalism 
based on the virial relations, see e.g. \citep{Osipkov2006}
for its formulation for the problem on hand.

The analytic results presented here verify, in an idealized limit, the those of some preliminary numerical simulations by 
\citet{Oh2000a}.  Those simulations were carried out for several clusters with more centrally-concentrated density 
distribution (i.e. a King model with C=1.8) embedded in one set of background potential (a King potential with C=0.5 
for a dwarf galaxy).  Follow-up numerical simulations are needed to verify that centrally concentrated clusters are 
more tightly bound by their self gravity and are more resilient to external tidal perturbation.
Although such
simulations have been carried for a galactic halo potential\citep{Oh1992a, Oh1995, Oh1992b}, 
follow-up investigations will be useful to explore the effects of tidal compression for 
centrally-condensed clusters subjected to orbit decay due to dynamical friction in 
a more general galactic potential. These investigations will be reported elsewhere.

\acknowledgments
The authors thank Gordon Ogilvie and John Papaloizou for support and hospitality
during the starting point of this project. We also thank the Department of Applied 
Mathematics and Theoretical Physics, Cambridge University and the Institute for 
Advanced Studies Tsinghua University for support.  
We also thank Avishai Dekel, Andrei Doroshkevich, Marla Geha, Puragra Guhathakurta, Anatoly Neishtadt,
John Papaloizou, Evgeny Polyachenko, Alexei Rastorguev, Alexander Polnarev, Ilya Shukhman and Scott Tremaine  for useful discussions and comments.


\bibliography{bib}{}
\bibliographystyle{aasjournal}

\end{document}